\documentclass[twocolumn,prl,10pt,superscriptaddress]{revtex4}
\usepackage{graphicx}
\usepackage{graphics}
\usepackage{amsmath}

\DeclareGraphicsExtensions{.eps}

\begin{document}

\title{A frequency and sensitivity tunable microresonator array for high-speed quantum processor readout} 

\author{J. D. Whittaker}\affiliation{D-Wave Systems, Inc., Burnaby, British Columbia, V5G 4M9, Canada}
\author{L. J. Swenson}\affiliation{D-Wave Systems, Inc., Burnaby, British Columbia, V5G 4M9, Canada}
\author{M. H. Volkmann}\affiliation{D-Wave Systems, Inc., Burnaby, British Columbia, V5G 4M9, Canada}
\author{P. Spear}\affiliation{D-Wave Systems, Inc., Burnaby, British Columbia, V5G 4M9, Canada}
\author{F. Altomare}\affiliation{D-Wave Systems, Inc., Burnaby, British Columbia, V5G 4M9, Canada}
\author{A. J. Berkley}\affiliation{D-Wave Systems, Inc., Burnaby, British Columbia, V5G 4M9, Canada}
\author{B. Bumble}\affiliation{Jet Propulsion Laboratory, California Institute of Technology, Pasadena, California 91109, USA}
\author{P. Bunyk}\affiliation{D-Wave Systems, Inc., Burnaby, British Columbia, V5G 4M9, Canada}
\author{P. K. Day}\affiliation{Jet Propulsion Laboratory, California Institute of Technology, Pasadena, California 91109, USA}
\author{B. H. Eom}\affiliation{Jet Propulsion Laboratory, California Institute of Technology, Pasadena, California 91109, USA}
\author{R. Harris}\affiliation{D-Wave Systems, Inc., Burnaby, British Columbia, V5G 4M9, Canada}
\author{J. P. Hilton}\affiliation{D-Wave Systems, Inc., Burnaby, British Columbia, V5G 4M9, Canada}
\author{E. Hoskinson}\affiliation{D-Wave Systems, Inc., Burnaby, British Columbia, V5G 4M9, Canada}
\author{M. W. Johnson}\affiliation{D-Wave Systems, Inc., Burnaby, British Columbia, V5G 4M9, Canada}
\author{A. Kleinsasser}\affiliation{Jet Propulsion Laboratory, California Institute of Technology, Pasadena, California 91109, USA}
\author{E. Ladizinsky}\affiliation{D-Wave Systems, Inc., Burnaby, British Columbia, V5G 4M9, Canada}
\author{T. Lanting}\affiliation{D-Wave Systems, Inc., Burnaby, British Columbia, V5G 4M9, Canada}
\author{T. Oh}\affiliation{D-Wave Systems, Inc., Burnaby, British Columbia, V5G 4M9, Canada}
\author{I. Perminov}\affiliation{D-Wave Systems, Inc., Burnaby, British Columbia, V5G 4M9, Canada}
\author{E. Tolkacheva}\affiliation{D-Wave Systems, Inc., Burnaby, British Columbia, V5G 4M9, Canada}
\author{J. Yao}\affiliation{D-Wave Systems, Inc., Burnaby, British Columbia, V5G 4M9, Canada}

\email{jwhittaker@dwavesys.com}

\begin{abstract}
  Superconducting microresonators have been successfully utilized as
  detection elements for a wide variety of applications.  With
  multiplexing factors exceeding 1,000 detectors per transmission
  line, they are the most scalable low-temperature detector technology
  demonstrated to date. For high-throughput applications, fewer
  detectors can be coupled to a single wire but utilize a larger
  per-detector bandwidth. For all existing designs, fluctuations in
  fabrication tolerances result in a non-uniform shift in resonance
  frequency and sensitivity, which ultimately limits the efficiency of
  bandwidth utilization. Here we present the design, implementation,
  and initial characterization of a superconducting microresonator
  readout integrating two tunable inductances per detector. We
  demonstrate that these tuning elements provide independent control
  of both the detector frequency and sensitivity, allowing us to
  maximize the transmission line bandwidth utilization. Finally we
  discuss the integration of these detectors in a multilayer
  fabrication stack for high-speed readout of the D-Wave quantum
  processor, highlighting the use of control and routing circuitry
  composed of single-flux-quantum loops to minimize the number of
  control wires at the lowest temperature stage.

\end{abstract}

\maketitle

\section{Introduction}

Since their first demonstration in 2003,\cite{day:2003}
superconducting microresonators have become increasingly popular for
low-temperature detector applications. They are an attractive option
due to their intrinsic frequency-domain multiplexing capability, the
wide availability of high-speed digital electronics, and breakthroughs
in material physics allowing very high sensitivities to be
achieved.\cite{leduc:2010, gao:2011, cecil:2012, devisser:2011,
  barends:2008, vissers:2012, catalano:2015} Arrays of these devices
have been successfully demonstrated at sub-mm,\cite{monfardini:2010,
  monfardini:2011, maloney:2010} far-infrared,\cite{swenson:2012,
  mckenney:2012} infrared/optical,\cite{mazin:2010}, and
x-ray\cite{quaranta:2013} wavelengths, for phonon
imaging\cite{swenson:2010, cruciani:2012, moore:2012} and in quantum
computing circuity.\cite{jerger:2011, chen:2012}.

Despite these achievements, superconducting
microresonator arrays typically suffer from inefficient bandwidth
utilization due to unavoidable fabrication imperfections.   As an example, the operating frequency $f$ of an LC resonator utilizing a parallel plate capacitor suitable for integration with a multi-layer fabrication stack will be sensitive to perturbations of the dielectric thickness $d$ according to the expression $\delta f/f = \delta d / 2 d$.  A realistic $\pm10\%$ tolerance in layer thickness for a modern fabrication facility results in up to $\pm$300 MHz shift for a nominal 6 GHz microresonator.  Global
variations affecting all devices uniformly can usually be compensated by a simple shift in the readout
center frequency, however, local variations are more problematic.
These cause each microresonator to exhibit a resonance frequency $f_0$
which deviates randomly from its expected value - often by more than a
linewidth.  The array designer is thus left with a choice of either
sparsely populating the available electronics bandwidth with
resonances such that stochastic resonance collisions are avoided, or
utilizing a high multiplexing factor and accepting a decreased array
yield due to overlapping resonances.

\begin{figure}[ht]   
\includegraphics[width=0.6\linewidth]{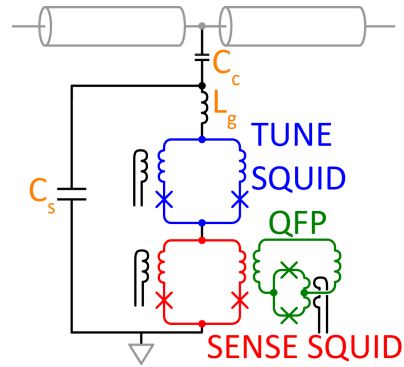}
  \caption{Schematic for a single FASTR detector.  As described in the
    text, the microresonator parameters are adjusted with local
    magnetic flux biases to the DC-SQUID loops which act as tunable
    nonlinear inductors.  Tuning these biases allows in-situ
    compensation for the unavoidable device variations inherent in
    fabrication.  While both DC-SQUIDs are tunable, only a single
    loop, labeled SENSE, is coupled to the measurement system.  For
    the D-Wave quantum processor, this is the last stage of a data
    shift register made from quantum flux parametrons
    (QFPs).}\label{figure-1}
\end{figure}

One way to make an array with efficient bandwidth utilization is to use superconducting microresonators whose frequency and sensitivity can be tuned in-situ, allowing compensation for local fabrication variations.  Tuning of these parameters allows an array of detectors to exhibit homogeneous sensitivity and uniform frequency spacing.  While superconducting microresonators with tunable frequencies have been previously demonstrated,\cite{mates:2008, whittaker:2014, vissers:2015, palacios-laloy:2008} this single adjustment knob is insufficient in the context of an efficiently packed array.   The detector sensitivity is equally important when the array is subject to a signal with a large dynamic range, as more sensitive detectors will exhibit a larger frequency shift than less sensitive detectors resulting in detector frequency collisions.  Adding a second tuning knob that also provides sensitivity tuning allows maximum packing efficiency to be achieved for a fixed electronic bandwidth.  A schematic of a Frequency And Sensitivity Tunable Resonator (FASTR) detector is shown in Fig.\ \ref{figure-1}. 

\begin{figure}[ht]
  \includegraphics[width=0.9\linewidth]{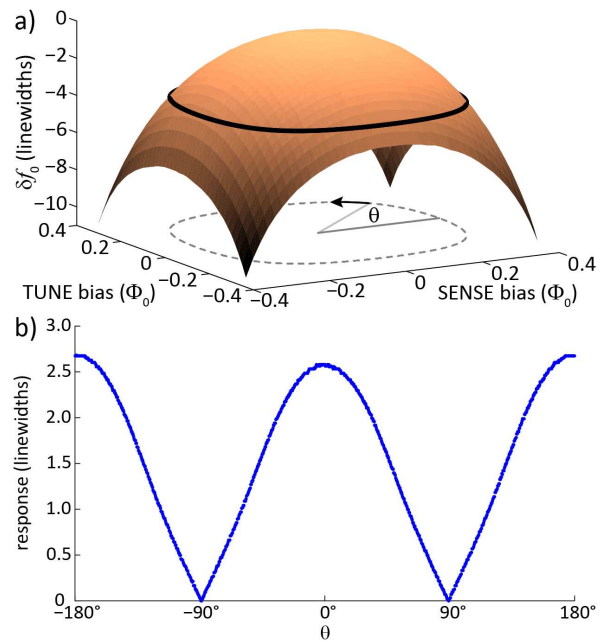}
  \caption{a) The measured resonance frequency of a prototype device
    to applied bias in the two DC-SQUID loops.  The unbiased resonance
    frequency of this device is $f_0 =$ 6.91 GHz.  The black line is a
    contour of constant frequency at $f_0=6.83$ GHz and the linewidth
    is 21 MHz. b) Response $\left|\delta f_0\right|$ in linewidths
    while moving along the contour.  The signal in this case was the
    flux change induced by the two states of the QFP magnetically
    coupled to the SENSE SQUID.  Note that for the low resonator
    quality factor $Q_r$ and large detector bandwidth discussed here,
    the noise for this readout is fixed and dominated by the readout
    electronics.  In this case, changes in responsivity directly
    impact the detector sensitivity.}\label{figure-2}
\end{figure}

The tuning parameters are realized using two DC-SQUID loops in series with a fixed geometric inductance $L_g$.  Each of these DC-SQUIDs are nonlinear inductances that can be adjusted using an external flux bias.  This inductive branch is in parallel with a shunt capacitor $C_s$.  External flux can be applied locally to each DC-SQUID loop either with individual analog wires, requiring $2N$ wires for an array of $N$ detectors, or with an array of flux DACs.\cite{bunyk:2014} This latter option can take advantage of an addressing network which requires only $O(\sqrt[3]{2n_{res}})$ wires, making control of even very large arrays straightforward.  Flux sensitivity is achieved by coupling one of the two loops to the system of interest.  Proper selection of the DC-SQUID biases is required to achieve a desired operating frequency and sensitivity.  The measured detector frequency of a prototype device (discussed below) for various flux biases in the two DC-SQUID loops is shown in Fig.\ \ref{figure-2}a. 

A contour of constant frequency has been drawn on the surface at a frequency shift of about $-$3.5 linewidths from the zero-flux resonance frequency.  Only the SENSE SQUID is coupled to the measurement system.  For the schematic shown in Fig.\ \ref{figure-1}, the FASTR detector is coupled to a quantum flux parametron (QFP) which is the last stage of a data shift register.  Modulating the state of the QFP while following this contour allows the device responsivity to be measured.  The result of this measurement is shown in Fig.\ \ref{figure-2}b.  Bias selection then consists of first choosing a contour of constant frequency followed by a traversal of the contour until the desired responsivity has been achieved.

Here we discuss the real-world example of utilizing a frequency-multiplexed array of FASTR detectors for readout of the D-Wave Two quantum processor.  We first briefly discuss the readout constraints imposed by the D-Wave Two quantum processor.  We then present a parametric design and physical realization of a single prototype FASTR detector suitable for processor integration.  While this specific application demonstrates a proof-of-principle of a frequency- and sensitivity-tunable superconducting microresonator, it should be kept in mind that this device can be broadly applied to a wide variety of magnetic-flux sensing applications including readout of the family of single flux quantum (SFQ) digital circuity.

\section{Readout constraints of the D-Wave Two quantum processor}

The current D-Wave quantum processor utilizes the quantum annealing algorithm to solve for the low lying energy states of the Ising spin Hamiltonian
\begin{equation}\label{isingHamiltonian}
 H = \sum\limits_{i} h_i \sigma_i^z + \sum\limits_{i<j}J_{ij}\sigma_i^z\sigma_j^z
\end{equation}
where the $\sigma_i$ are Pauli matrices for the $i$th spin, $h_i$ are local energy scaling factors, and the non-zero coupling terms $J_{ij}$ reflect the underlying connectivity of the processor graph.  For the D-Wave Two processor, the graph consists of 64 repeated 8 qubit unit cells.  Within a cell the qubits are connected in a K4,4 planar topology.  Each qubit is further connected to two qubits in neighboring unit cells to yield 6 connections per processor qubit. The processor topology, qubit parametric design, problem specification (consisting of setting the $h_i$ and $J_{ij}$ terms in Eq.\ \ref{isingHamiltonian}), and the annealing algorithm have been previously described.\cite{harris:2010, harris:2010b, johnson:2011, bunyk:2014} It is important to note that during the annealing procedure qubit entanglement\cite{lanting-ent:2014} is eventually frozen out and all qubits are projected into a classical flux state before readout.

\begin{figure}[ht]
  \includegraphics[width=0.7\linewidth]{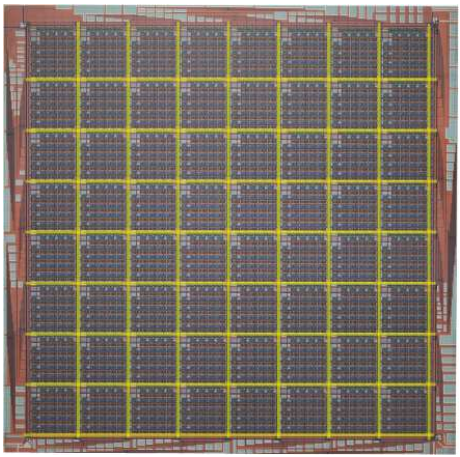}
  \caption{Picture of the D-Wave Two processor chip, made up of 64
    eight-qubit unit cells. The shift register streets are highlighted
    yellow and lead to the outside of the processor where it is
    straightforward to locate FASTR detectors.}\label{figure-3}
\end{figure}

\begin{figure}[ht]
  \includegraphics[width=0.9\linewidth]{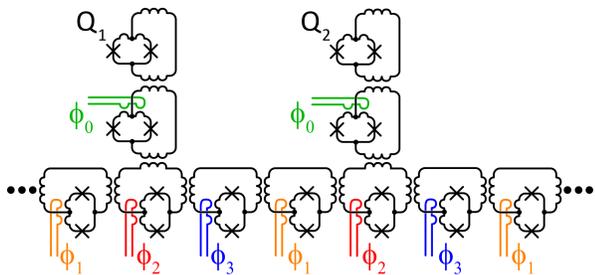}
  \caption{Simplified schematic of a portion of the QFP-based shift
    register. $Q_1$ and $Q_2$ are qubits and $\phi_{0...3}$ are flux
    biases that control data flow.}\label{figure-4}
\end{figure}

Processor readout entails detecting the flux state of all qubits at
the end of a computation sequence.  The readout architecture consists
of a shift register lattice interlaced between the 8-qubit unit
cells. A picture of the processor chip, with the shift register
lattice highlighted in yellow, is shown in Fig.\ \ref{figure-3}.  The
shift register eliminates the need for an individual detector per
qubit and is used to move state information from individual qubits out
to detectors on the chip perimeter.  A simplified portion of the shift
register is shown schematically in Fig.\ \ref{figure-4}.

A shift register stage is composed of a QFP that can be modulated between a monostable unlatched and a bistable latched state using a local flux bias $\phi_n$.  In the latched state, information is stored in the QFP as either circulating or counter-circulating current.  This current is a flux source which applies a flux bias to the neighboring shift register stages held in the unlatched state.  The next stage in the forward direction is subsequently latched before the source stage is unlatched allowing information to be passed down the shift register.  Note that two unlatched stages are required between every latched stage in order to protect against back-action from data further ahead in the shift register.  As indicated in Fig.\ \ref{figure-4}, three QFP flux biases $\phi_{1...3}$ are thus required to pass information along the shift register.  A separate QFP stage, biased by $\phi_0$, is directly connected to each qubit and is used to selectively copy data from the qubit to the rest of the shift register.

The topology and operating speed of the shift register set crucial constraints for the FASTR readout array.  For the processor shown in Fig.\ \ref{figure-3}, there are 8 vertical and 8 horizontal shift registers.  As data can be moved in either direction along the shift register, there are 32 points where detectors can easily be placed. Note, however, that the shift register can only move in one direction at a time so that only half of these detectors can be operated simultaneously.  The detector on the other end of each line provides redundancy for shift register breaks caused by inoperable QFPs. While in principle it is possible to increase the shift register fan-out by utilizing more stages, this option increases circuit complexity.

\begin{figure}[ht]
  \includegraphics[width=0.8\linewidth]{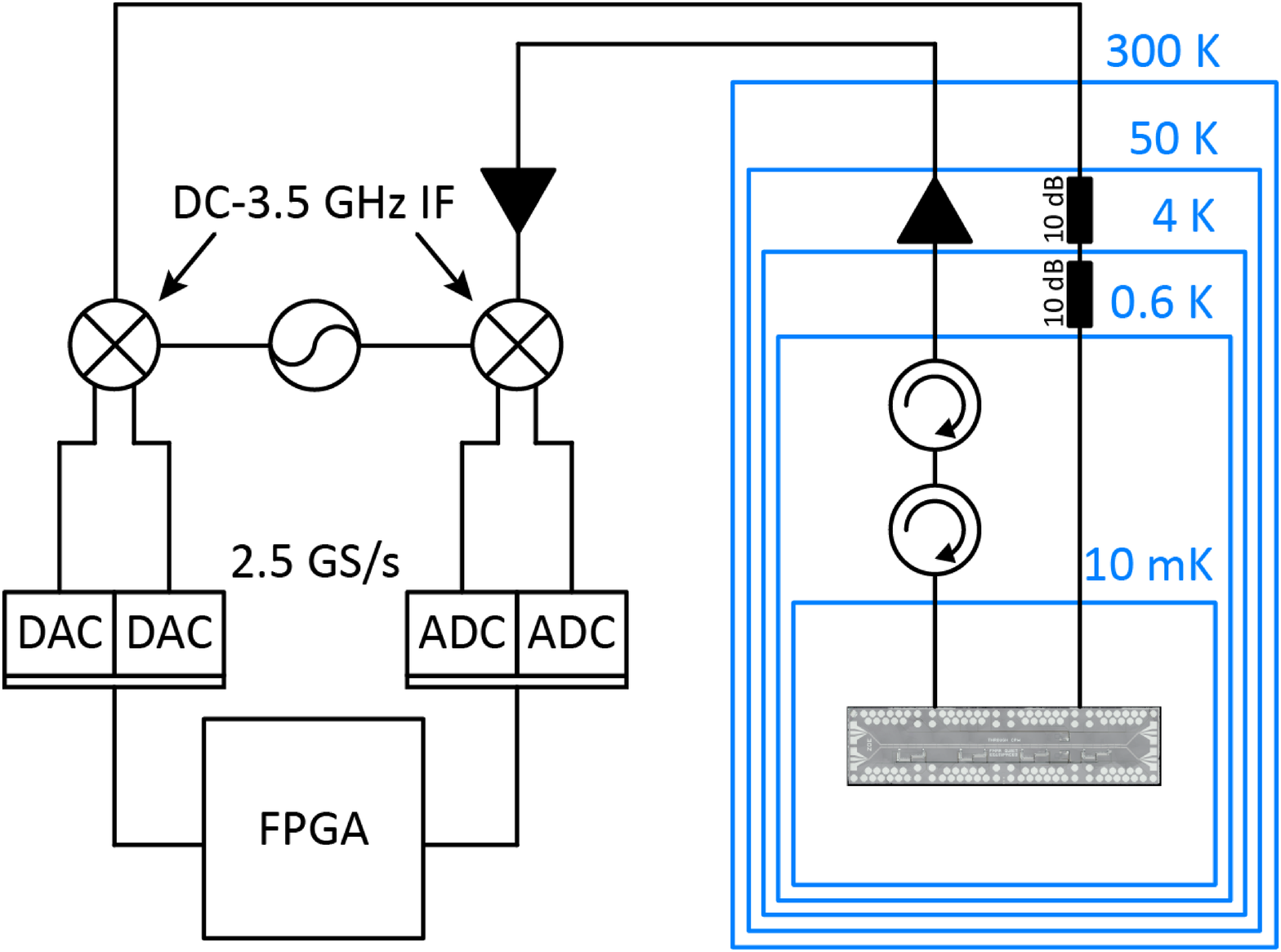}
  \caption{Simplified system electronics schematic for FASTR array
    readout.  The digital-to-analog and analog-to-digital converters
    are phase locked and sampled at 2.5 GSPS.}\label{figure-5}
\end{figure}

The shift register operation speed is set by the bandwidth of the QFP flux bias lines $\phi_n$.  Currently these lines utilize a 30 MHz low-pass filter to limit noise.  As the shift register is composed of three biases $\phi_{1...3}$, each shift register line can deliver data at $\sim$10 Mbits/s.  If only a single detector is attached to each shift register line, the detector bandwidth should match or somewhat exceed 10 MHz.

Although the attainable readout speed for a FASTR detector is set by the microresonator linewidth, more than a single linewidth of electronic bandwidth is required per detector.  Additional bandwidth is required to accommodate shifting of the resonance frequency due to modulation of the signal coupled into the SENSE SQUID.  Even more electronic bandwidth is required to avoid electronic crosstalk with neighboring microresonators, as the resonance appreciably perturbs the microwave transmission beyond the nominal linewidth.

Given the preceding constraints, the FASTR readout for the D-Wave Two quantum processor is designed to utilize resonances with 19.5 MHz linewidths, modulated by 1 linewidth by the shift register data.  Each detector is allotted 4 linewidths of electronic bandwidth to minimize crosstalk.  A total of 2.5 GHz of readout electronic bandwidth is therefore required to readout all 32 detectors on a single transmission line.  Considering this, the current D-Wave FASTR readout is designed to operate using a 2.5 GHz bandwidth with an array center frequency of 6 GHz.  A simplified schematic of the readout circuit is shown in Fig.\ \ref{figure-5}.

The readout design is similar to Kinetic Inductance Detector (KID) based readouts with the notable feature of a large 2.5 GSPS sampling frequency of the baseband digitizing electronics.\cite{bourrion:2011, yates:2009, mchugh:2012}  In contrast to KID readouts used for imaging which continuously interrogate the microresonator array, the FASTR readout only measures the array state after completion of a shift register operation cycle.  The $\sim$35$\%$ duty cycle reduces the data rate to a level at which the Field Programmable Gate Array (FPGA) used for signal processing can continuously sustain.  As previously mentioned, the center frequency of the 2.5 GHz of readout bandwidth should be adjustable in order to compensate for any global shift in the detector frequencies.  For the current system, global shifts up to $\pm$750 MHz can be tolerated by adjusting the local oscillator frequency.  This limit is imposed by the 4-8 GHz operating band of the commercial microwave components used in the readout chain.

Other constraints are imposed when integrating FASTR devices (Fig.\ \ref{figure-1}) into the D-Wave Two quantum processor, which utilizes a multilayer fabrication stack consisting of 6 niobium metal layers separated by planarized dielectric layers. A particular challenge is to realize FASTR devices with this multilayer stack that achieve sufficiently high intrinsic quality factors $Q_i$. The linewidth requirement discussed previously dictates that the microresonator quality factors $Q_r$ are of order $f_0/\Delta{f}=6\mathrm{e}9/19.5\mathrm{e}6\sim300$.  In order achieve full signal modulation and minimize processor heating due to dissipation by the FASTR readout, it is desirable for $Q_i$ to be as high as possible.  Achieving the limit $Q_i \gg Q_r$ has the added advantage that $Q_r$ is then simply determined by the coupling to the microwave circuit and can be easily set in design with the coupling capacitance $C_c$.

The primary source of dissipation in superconducting microresonators has been extensively studied,\cite{martinis:2005, oconnell:2008, gao:2008b} and low-loss microresonators made of interdigitated capacitors on crystalline substrates routinely achieve intrinsic quality factors exceeding a million.\cite{leduc:2010, geerlings:2012, chang:2013}  This low loss is achieved by engineering resonators with  the electric field energy constrained to vacuum or the underlying low-loss crystalline substrate.  The surface, where lossy dangling bonds and oxides are present, are specifically avoided in these devices.\cite{gao:2008c, quintana:2014, dial:2015}  For example, one technique that is commonly used to improve the $Q_i$ of planar interdigitated capacitor is to increase the finger spacing which decreases the participation ratio of the surface defects.

Compared with single-layer microresonator designs on a crystalline substrate, microresonators incorporating deposited dielectrics suitable for integration in a multilayer fabrication stack typically exhibit much higher loss due to the presence of two-level systems (TLS) in the amorphous dielectric.\cite{martinis:2014}  In this case, simple geometric changes such as increasing the finger spacing of an interdigitated capacitor do not result in an improved device $Q_i$ as the electric field is displaced but remains within the lossy deposited dielectric.  Fortunately, these dissipation pathways become saturated above a critical electric field, such that $Q_i$ increases with drive power.  In this case, by minimizing the dielectric volume used for fabricating the resonator dielectric, saturation can be achieved for a minimal stored resonator energy.  This allows high intrinsic quality factors and minimal heating to be achieved for modest microwave drive powers.

Considering the need to minimize the volume of lossy dielectric, a parallel plate geometry with a minimal dielectric thickness is particularly suitable for the integrated FASTR design.  Finally, a lumped-element geometry utilizing a parallel-plate capacitor comes with the added advantage that, as opposed to distributed resonator geometries, no harmonics are inherent in the design.  This allows the readout to occupy more than a single octave of bandwidth which is an important consideration for scaling to large array sizes.

\section{Characterization of a prototype device}

\begin{figure}[ht]
  \includegraphics[width=1\linewidth]{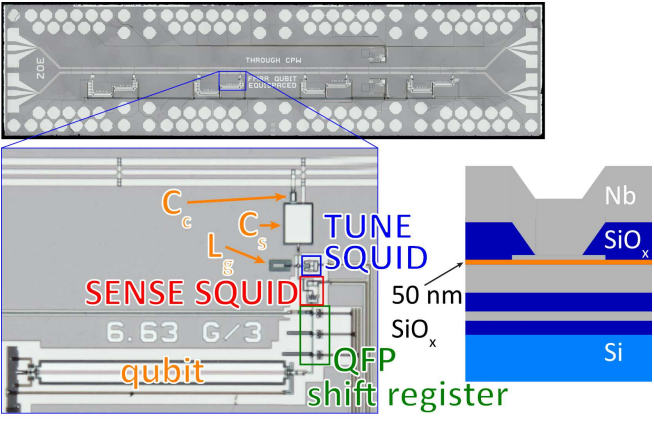}
  \caption{Prototype FASTR detector with three QFP shift register
    stages, designed to have $C_s=1.7$ pF, $C_c=70$ fF, $Q_c=338$,
    $L_g=324$ pH, $f_0=6.628$ GHz, and Josephson junction $I_c=11$
    $\mu$A. The fabrication stack is also shown, highlighting the 50
    nm thick capacitor dielectric layer.}\label{figure-6}
\end{figure}

We have designed, fabricated, and measured a prototype FASTR detector suitable for integration with the D-Wave Two quantum processor.  The layout and fabrication stack are shown in Fig.\ \ref{figure-6}.  Testing was performed at 10 mK.  The detector resonance frequency with zero flux in either SQUID loop was measured to be 6.91 GHz.  This is a large but still acceptable 300 MHz above the design value and attributable to fabrication variation in the capacitor dielectric layer thickness.  While only a single prototype FASTR detector was characterized in this initial measurement, the multiplexed readout system shown in Fig.\ \ref{figure-5} was used for electrical readout while operating at the full 2.5 GSPS data rate.

\begin{figure}[ht]
  \includegraphics[width=1\linewidth]{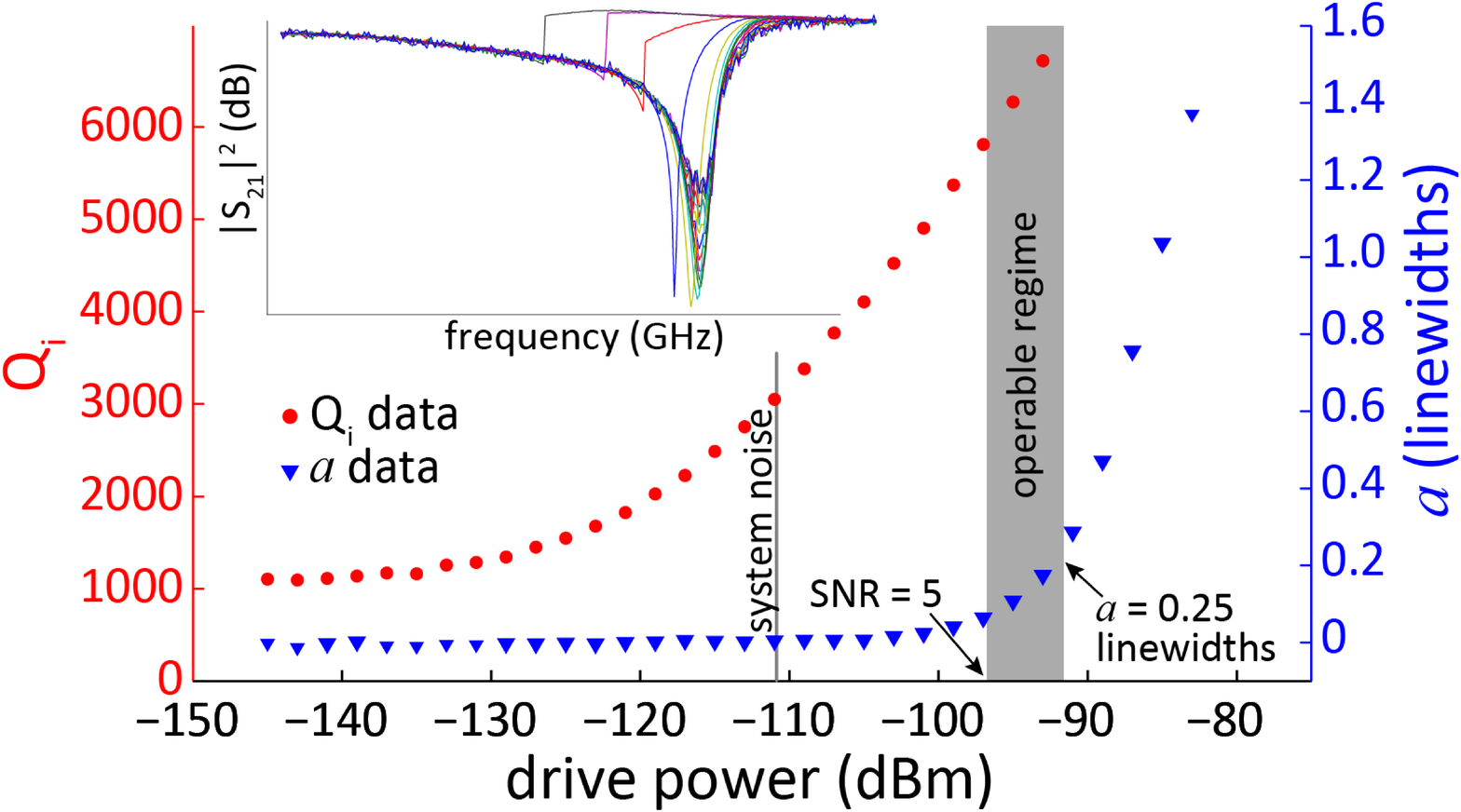}
  \caption{Resonator intrinsic quality factor $Q_i$ and nonlinear
    shift $a$ as functions of the drive power, under typical bias
    conditions at $f_0=6.84$ GHz. These parameters have been extracted
    from the data shown in the inset.  The resonator should be
    operated at a sufficient power to provide a large amplitude SNR
    required for high fidelity readout while avoiding intrinsic device
    nonlinearity.  For the prototype FASTR detector, the region with
    an SNR $>$ 5 and weak device nonlinearity ($a$ $<$ 0.25) is
    indicated as the operable region. The coupling quality factor
    $Q_c$ for this device was 329 and has an unbiased resonance
    frequency of $f_0=6.910$ GHz.}\label{figure-7}
\end{figure}

The fabricated prototype device featured a parallel plate capacitor
with a nominally 50 nm thick sputtered silicon oxide
dielectric. $S_{21}$ transmission data were taken and fit to the
expression for a transmission line shunted by a parallel RLC resonator
\begin{equation}
  S_{21}=1 - \frac{Q_r}{Q_c}\frac{1}{1+2iQ_rx}\label{lorentzian-fit}
\end{equation}
where $x = \left(f-f_0\right)/f_0$ is the fractional detuning from
resonance.  The results are shown in Fig.\ \ref{figure-7} for a
variety of powers.  The low-power intrinsic quality factor of the
resonator with this capacitor was measured to be $Q_i\sim1000$. The
fit results show $Q_i$ changing with power (red dots), exhibiting
two-level system (TLS) saturation behavior with a saturation field of
$|\vec{E}| = 50$ V/m.  While the low-power $Q_i$ is not much greater
than the $Q_r \sim 300$ required for the 32-resonator design, above
the saturation threshold $Q_i$ increases dramatically.

\begin{figure}[ht]
  \includegraphics[width=1\linewidth]{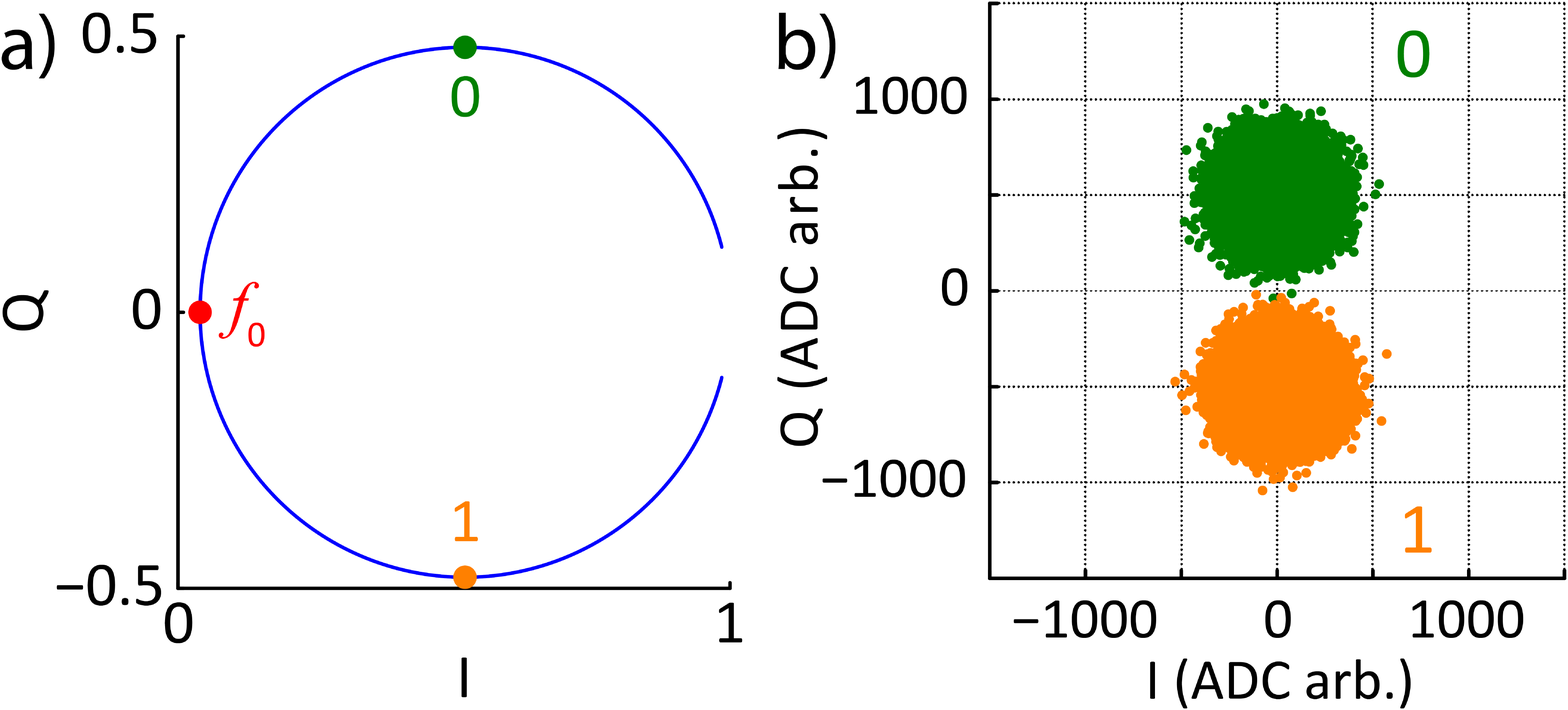}
  \caption{a) The two QFP states, 0 and 1, shown at $\pm$1/2 linewidth
    from the FASTR resonance $f_0$, which is achieved when both biases
    are properly tuned. b) Complex plane plot of prepared qubit states
    read out by the FASTR system, at a SNR$\sim$5. The data have been
    translated and rotated for simple state discrimination along
    $Q=0$.}\label{figure-8}
\end{figure}

Fig.\ \ref{figure-7} also shows how the resonance frequency changes with drive power (blue triangles) in units of linewidths according to $a=\left(f_{0}'-f_0\right)Q_r/f_0'$, where $f_{0}'$ is the low-power resonance frequency.  This non-linear response to stored energy is characterized by the Duffing oscillator nonlinearity parameter $a=\left(\alpha{Q_r}/4\right)\left(I/I_c\right)^2$.\cite{zmuidzinas:2012, swenson:2013} Here, $\alpha=L_J/(L_g+L_J)$ characterizes the contribution of the Josephson Junction kinetic inductance $L_J = \Phi_0/2 \pi I_c$ to the total inductance, $I$ is the current through the resonator, and bifurcation occurs at $a \sim 0.77$ linewidths.  Keeping the microwave drive power sufficiently low assures only minimal nonlinearity ($a<0.25$ linewidths) and well-behaved detector operation.  This requirement sets the upper bound to the operable region labeled in Fig.\ \ref{figure-7}. The lower bound is set by the a minimum desired amplitude SNR of 5 which gives a bit error probability less than $10^{-6}$.  We have measured the system noise temperature to be 7.9 K, limited primarily by the first stage amplifier and the insertion loss of the cryogenic isolators used to prevent amplifier back-action from disturbing the processor.  The microwave drive power should exceed $P_g=$ -96 dBm to achieve the desired SNR assuming the tone is fully modulated by the 1 linewidth resonance shift.  Full modulation is achieved by requiring $Q_i \gg Q_c$ and biasing symmetrically around the resonance, which maps the two qubit states onto the resonance curve as shown in Figure\ \ref{figure-8}a.

\section{FASTR readout of a qubit state and noise}

As a proof-of-principle of full-system operation, the prototype FASTR detector was used to readout a qubit via the three-stage shift register shown in Fig.\ \ref{figure-6}. The detector was driven at $P_g=-98$ dBm ($a=0.05$ linewidths), slightly below the SNR = 5 lower limit of the operable region indicated in Fig.\ \ref{figure-7}.  A large positive flux bias applied to the qubit, followed by annealing, allows for preparation in a known state.  Using a large negative flux bias allows the qubit to be prepared in the opposite state.  After qubit annealing, the resulting state can be passed along the shift register and read out with the FASTR detector. Repeating this procedure many times with a known data pattern allows the readout fidelity to be assessed.  A plot of the measured fidelity data is shown in Fig.\ \ref{figure-8}b, where a linear transformation to the complex transmission has been applied to simplify state discrimination.  This process of transformation and discrimination was applied to all subsequent qubit measurements.  From the fidelity data, a bit error probability of $\sim10^{-5}$ was calculated, which matches well with expectation for the device parameters and the microwave drive power utilized.  Note that while an SNR $<$ 5 is sufficient for verification of this prototype device, the measurement power will be properly adjusted before full processor readout to ensure the desired bit error probability of $<10^{-6}$ is achieved.

\begin{figure}[ht]
  \includegraphics[width=1\linewidth]{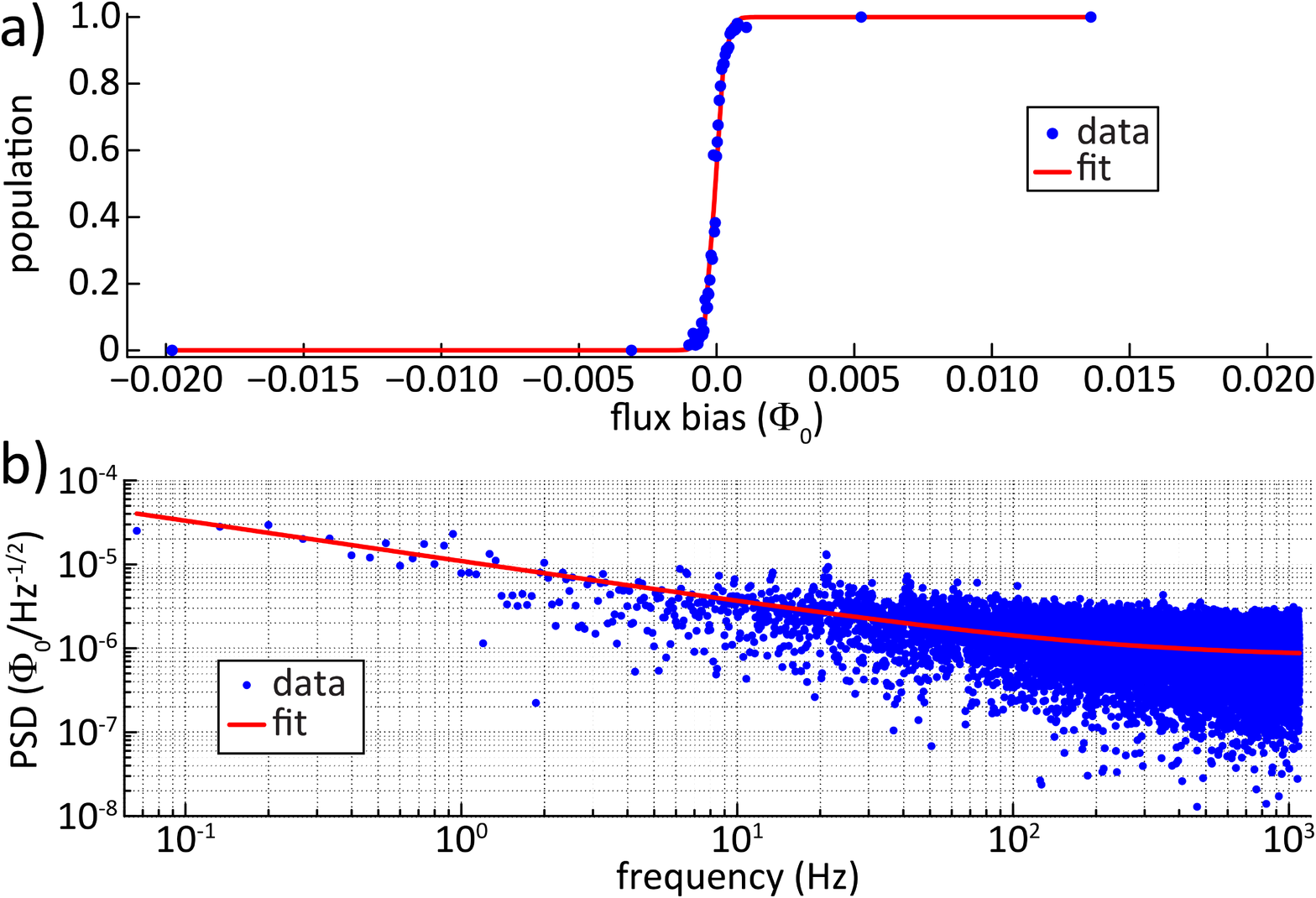}
  \caption{a) Qubit population measurement, read out with a FASTR, as
    qubit flux bias was swept. b) Noise PSD measurement and fit at a
    sampling rate $1/\tau_s=1/3.6\mu{s}$. The $1/f$ noise amplitude at
    1 Hz was 11 $\mu\Phi_0/\sqrt{Hz}$, while the white noise was 0.8
    $\mu\Phi_0/\sqrt{Hz}$.}\label{figure-9}
\end{figure}

The previous measurements validate the operation of a FASTR detector and the readout system.  It is now possible to repeat two of D-Wave's standard qubit characterization measurements: a qubit population measurement and a noise power spectral measurement (PSD).\cite{lanting:2009,lanting:2014}  The qubit population measurement was made while sweeping the qubit flux bias through degeneracy, with the results shown in Fig.\ \ref{figure-9}a.  The fit was done with the equation
\begin{equation}
P=\frac{1}{2}\left[1+\tanh\left(\frac{\Phi_x}{2W}\right)\right],
\end{equation}
where $\Phi_x$ is the externally applied flux and $2W$ is the width of the transition. Inverting this fit gives a mapping between qubit population and $\Phi_x$.  When annealed at a fixed bias point near the center of the transition, on average the qubit population should always return the same value.  However, noise will cause the population to wander.  Repeated measurement of the population and conversion into an equivalent external flux allows a noise PSD to be calculated. The result of this measurement using the prototype FASTR detector is shown in Fig.\ \ref{figure-9}b.  The noise spectra show a combination of $1/f$ noise and white noise $w_n$, where $w_n=\tau_s4W^2$.  The $1/f$ noise component is due to intrinsic device noise, and the white noise is due to the statistical nature of the qubit measurement and drops with increased sample rate or a narrower transition width.  Note that traditional DC-SQUIDs are dissipative.  Fast operation of these devices results in an increased sample temperature and consequently a wider transition width $W$.  This heating can be reduced by increasing the time per sample $\tau_s$.  However in either case $w_n$ is increased.\cite{lanting:2009,lanting:2014}  Thus the low dissipation and fast duty cycle allowed by a FASTR detector results in a low white noise level as compared with a DC SQUID readout.

\section{Perspective on readout scaling}

Scaling of the FASTR readout is constrained by the achievable device $Q_i$, available readout bandwidth, and the number of analog wires required to tune the individual resonance frequencies.  Based on commercially available electronics, the FASTR array was designed for a 2.5 GHz band centered around 6 GHz. Placing resonators at the ends of each shift register street, the number of resonators for the D-Wave processor goes as $n_{res}=4\sqrt{n_{cells}}$.   Table\ \ref{table-scaling} shows the scaling of the per-resonator bandwidth and required device $Q_i$ for a fixed 2.5 GHz bandwidth readout design.
\begin{table}[h]
  \begin{tabular}{| c |c | c | c | c | c | c |} \hline
    \textbf{$N_{qubits}$} & \textbf{$n_{cells}$} & \textbf{$n_{res}$} & \textbf{$\Delta f$ (MHz)} & \textbf{$Q_c$} & \textbf{$Q_i$} & \textbf{$N$} \\ \hline
    512  & 64  & 32 & 19.5 & $\sim$240-370  & $>$3,700 & 4 \\ \hline
    1152 & 144 & 48 & 13.0 & $\sim$360-560  & $>$5,600 & 5 \\ \hline
    2048 & 256 & 64 & 9.8  & $\sim$490-740  & $>$7,400 & 6 \\ \hline
    3200 & 400 & 80 & 7.8  & $\sim$610-930  & $>$9,300 & 6 \\ \hline
    4608 & 576 & 96 & 6.5  & $\sim$730-1110 & $>$11,100 & 6 \\ \hline
  \end{tabular}
  \caption{Scaling of the FASTR readout scheme with the size of processor.  $N$ is the number of additional analog wires required as discussed in the text.   The stated $Q_i$ assumes the need to meet the condition $Q_i \gg Q_c$ to minimize chip heating and design effort.}\label{table-scaling}
\end{table}
Figure\ \ref{figure-7} shows that for the SiO$_x$ dielectric used for the prototype device, a $Q_i$ of $\sim$6,000 was
achieved in the operating regime.  Increasing the number of devices per line beyond $\sim$48 thus requires incorporation of lower loss deposited dielectrics such as hydrogen-rich amorphous silicon.\cite{mazin:2010b}

A straightforward way to scale the readout is to increase the readout bandwidth.  Already, digitizing components with bandwidths exceeding 5 GHz and sufficient dynamic range are commercially available.  As the design here utilizes a parallel-plate capacitor incorporated in a lumped-element resonator geometry which does not exhibit harmonics, more than one octave of bandwidth can be readily utilized.  Perhaps the greatest difficulty for implementing large bandwidth electronics is the very high data rates that must be handled by the signal-processing hardware.  Fortunately, very large FPGAs are becoming available that should be sufficient for this task.

Each FASTR device requires two external flux biases.  $2n_{res}$ lines will be needed if each of these are supplied with individual analog lines. A 32-resonator system would require already a somewhat impractical 64 analog lines.  As a better alternative, the D-Wave Two processor uses superconducting digital-to-analog flux converters to apply flux biases to thousands of devices though an addressing scheme that scales as $O(\sqrt[3]{2n_{res}})$.\cite{bunyk:2014}  For instance, a 96-resonator system would require the addition of only 6 analog lines using this addressing scheme, which is far more reasonable than the 192 lines a per-device addressing scheme would require.

\section{Conclusion}

We have demonstrated a frequency- and sensitivity-tunable resonator detector suitable for a wide variety of magnetic flux sensing applications.  The detector tuning parameters allow for compensation of detector-scale fabrication variations, while shifting the center frequency of the readout electronics allows for compensation of wafer-scale variations.  Using a combination of these two techniques, the resonator frequencies can be spaced uniformly and a homogenized flux-sensitivity can be realized.  In this way a maximum packing efficiency can be achieved for a fixed electronic bandwidth.  The utility of this device includes readout of the D-Wave quantum processor, SFQ digital circuitry, and indeed any magnetic flux sensing applications requiring arrays of densely multiplexed detectors.

\section{Acknowledgements}

The authors would like to thank R. Neufeld and D. Walliman for
providing photographs of samples, as well as C. Enderud, C. Baron, and
M. Babcock for sample preparation and cryogenic support. A portion of
this research was carried out at the Jet Propulsion Laboratory,
California Institute of Technology, under a contract with the National
Aeronautics and Space Administration.


\end{document}